%% file: main-arxiv.tex
\documentclass[11pt]{article}

\usepackage{fullpage}

\usepackage{graphicx}
\usepackage{setspace}
\usepackage{xcolor}
\usepackage{bbm}
\usepackage{amsmath}
\usepackage{bbm}
\usepackage{url}

\newcommand{\reals}{\mathbbm{R}}

\newcommand{\mI} {\ensuremath{\mathcal{I}}}
\newcommand{\mJ} {\ensuremath{\mathcal{J}}}
\usepackage{nicefrac}
\newcommand{\rbr}[1]{\left(\,#1\,\right)}

\renewcommand{\paragraph}[2]{\vspace{2mm}\noindent \textbf{#2.}}

\title{Generative AI as Economic Agents%
\footnote{We would like to thank Nageeb Ali, Drew Fudenberg, John Horton, Adam Kalai, David Kempe, Akshay Krishnamurthy, Kevin Leyton-Brown, Yishay Mansour, Sam Taggart, and Clayton Thomas for their many helpful comments.
}}

\author{Nicole Immorlica%
\footnote{Microsoft Research New England,  
\texttt{\{nicimm, brlucier\}@microsoft.com}.}
\and Brendan Lucier\footnotemark[1] 
\and Aleksandrs Slivkins%
\footnote{Microsoft Research New York City,
\texttt{slivkins@microsoft.com}.}
}

\date{May 2024}

\begin{document}

\maketitle

\begin{abstract}
Traditionally, AI has been modeled within economics as a technology that impacts payoffs by reducing costs or refining information for human agents.
Our position is that, in light of recent advances in generative AI, it is increasingly useful to model AI itself as an economic agent.
In our framework, each user is augmented with an AI agent and can consult the AI prior to taking actions in a game.  The AI agent and the user have potentially different information and preferences over the communication, which can result in equilibria that are qualitatively different than in settings without AI.
\end{abstract}

\input{main-body}

\begin{small}
\bibliographystyle{plain}
\bibliography{refs.bib,bib-abbrv.bib,bib-LLMs.bib}
\end{small}

\appendix

\input{main-appendix}

\end{document}

%% file: main-body.tex
In economic theory models, agents take potentially costly actions to maximize their expected utility given their information about the world and others' strategies.  In some models, agents have access to technologies that help by reducing action costs or providing informative signals.
Researchers have tended to model AI as such a technology.
However, given the recent advancements in generative AI, and particularly Large Language Models (LLMs), we claim this new technology is itself sometimes best modeled as an economic agent.

Why do we suggest modeling generative AI as an agent in a game?  After all, it is just a type of machine-learning model
that predicts the next token in a sequence or samples content from a learned distribution.
A notable aspect of
generative AI, however, is its ability to generate novel content based on an implicit but vast ``common-sense'' understanding of the world.%
\footnote{
While such anthropomorphic metaphors can be misleading, our argument requires only a \emph{semblance} of common-sense reasoning.
} As such, generative AI models (and systems built upon them) can be leveraged as virtual consultants
that assist, analyze, or even strategize on behalf of their users.
We argue below that, much like a human consultant, an AI-powered virtual consultant exhibits features typical of an economic agent.

Indeed, like a human consultant, an AI consultant has an \emph{information set}, or a collection of knowledge and beliefs about the world and other actors therein.  This information can come from the training data, prompts, or external sources. Trained on massive datasets, AI consultants may have different, possibly much richer, signals about the world than any human.
{Moreover, the potential to learn from interactions with their entire user base opens up opportunities to adapt and integrate new knowledge more quickly than a human consultant.}

An AI consultant also has an \emph{action space} consisting of outputs to queries.  Similar to a human consultant, this action space can be very general, including, e.g., natural language
or source code.
However, current social norms impose constraints on the environment in which the AI can take these actions.  Namely, the AI is constrained to a virtual environment: its communication with a user.  Thus, AI has \emph{limited agency}; while it has agency over its responses, the user retains the power to make (or, equivalently, to veto) any payoff-relevant actions, possibly based on its interactions with the AI.\footnote{
We model AI in this way to reflect our focus on its  role as a \emph{consultant}. Our model also applies when AIs can directly take actions in the real world,
as long as the user retains veto power over such actions.
 }

Finally, an AI consultant has objectives and constraints ingrained during training, fine-tuning, and orchestration.
This induces the AI to act as though it is maximizing some implicit \emph{preferences}.%
\footnote{We distinguish between ``inherent" preferences of the AI and those induced by
user prompts (e.g., ``you are a candy-loving baby"). Note that the former can shape the latter by, e.g., influencing the manner and extent to which prompts are followed.
}
Importantly, similar to human consultants, these preferences may not be perfectly aligned with the user's preferences and are not directly under the user's control. However, unlike human consultants, the AI, once deployed, has a limited view of the world that typically  consists only of communications with its user.  Thus its preferences can only be a function of this communication transcript rather than real-world outcomes.%
\footnote{For example, a human consultant paid to work on one part of a larger project might prefer that the project is ultimately successful because this would look good on a resume. In contrast, an AI consultant in our model has no awareness of (and thus no incentives regarding) a project's real-world outcome beyond the user's instructions.}

Thus, our view is that AI-based virtual consultants can be reasonably modeled as economic agents.
{As discussed above, they have limited agency relative to
human consultants, as well as limited worldview and preferences, but can potentially be more knowledgeable and adaptive}.
This view of AI as an economic agent includes, as a special case, classic models that treat AI as a technology that directly reduces action costs or provides informative signals, as these can be viewed as agents with trivial preferences and particularly structured information and action sets.

AI consultants can also be drastically cheaper than human
consultants, and hence applicable in a wider array of contexts.
Across these contexts, they can play three (overlapping) roles:
\emph{assistant}, completing specific tasks;
\emph{analyst}, revealing and conveying information from built-in knowledge or external sources;
and \emph{strategist}, proposing actions for a user and possibly
accounting for other parties’ reactions. The three roles can lead to different modeling choices from an economic perspective.


Synthesizing these considerations, we propose augmenting economic theory models so that traditional agents (henceforth \emph{users}) are endowed with AI-based consultants (henceforth \emph{AI agents} or simply \emph{AIs}) and can interact with the latter before taking actions.
In what follows, we put forward a
formal framework to study the impact of AI agents as a step toward informing their design and deployment.
We instantiate this framework with several examples (including examples where preferences are misaligned), outline potential research questions, and relate our framework to existing literatures in economics and AI.  {Our view is that tools and techniques from the theory of economics and computation are perfectly suited to the challenges that arise in a world of AI agents, where the inherent features of AI systems meet the game-theoretic implications of individual preferences and multi-agent interactions.}

\section{A General Model}

\paragraph*{Baseline Model: Single-Shot Game without AI Agents}
The baseline for our model is the following general game-theoretic setup.
There are $n$ (human) players who interact in a game.  Each player $i$ has a (finite or infinite) action space $A_i$.  An action profile $\vec{a} \in A := \prod_i A_i$ denotes a choice of action by each player.  There is also a special agent called Nature who selects a state of nature $\omega \in \Omega$ at random from a commonly-known distribution. Each player $i$ is endowed with an information structure: a partition $\mI_i$ of $\Omega$.\footnote{That is, $\mI_i$ is a collection of disjoint subsets whose union is $\Omega$.}

The game proceeds as follows.
Nature first selects a state of nature $\omega$.  Each player $i$ is then endowed with an information set $I_i \subseteq \Omega$ such that $\omega \in I_i\in \mI_i$, which describes her (partial) information about the chosen state of nature. Players then simultaneously select actions.  Finally, each player receives a payoff $u_i(\vec{a}, \omega)$ that depends on the actions of all players and the state of nature.

This model can capture multiple game-theoretic scenarios and could be paired with various solution concepts.  If there is only a single state of nature, this is a standard simultaneous-move game, and one might consider pure or mixed Nash equilibria.  To give another example, the state of nature might consist of a vector of types, one per human player, and Nature might draw this state from a known product distribution.  Each player's information set might then reveal their own type, but not that of the other players.  Then we are in a Bayesian model of incomplete information, and we may be interested in its Bayes-Nash equilibrium.  Our framework can also be applied to other game formats such as extensive form games in which the information sets of the players contain (possibly incomplete) information about the history of play.

\paragraph*{AI Agents}
We augment our baseline model by associating an AI agent with each human player.  The human and the AI agent interact through a communication protocol, in which they alternate sending messages from a space $M$ of potential messages beginning with the human player.  Communication can proceed in multiple rounds, resulting in a \emph{transcript} $\tau_i$ (i.e., an ordered sequence of messages) between human player $i$ and their AI agent.  Our only assumption about the messaging protocol is that the human player can always choose not to send a message, effectively terminating the communication procedure.  Furthermore, and central to our work, the AI agent's payoff might be misaligned with that of its human player.

Like the human players, each AI agent $i$ has an information structure $\mJ_i$ and an information set $J_i\subset \Omega$ such that $\omega\in J_i \in \mJ_i$ that specifies what they know about the state of nature.  Each AI agent $i$ also has a payoff $v_i(\tau_i, \omega)$ that can depend on the state of nature and their transcript, but is otherwise independent of the actions chosen by the human players.  The information set held by an AI agent is not necessarily the same as the information set of their human user, and indeed a key aspect of AI agents is that they may have access to information that their associated human player does not.

We incorporate AI agents into the timing of our general game as follows.  Nature first selects a state $\omega \in \Omega$, as before.  Each human player $i$ then engages in communication with their associated AI agent, generating a (possibly empty) message transcript $\tau_i$.  There is a cost function $c_i$ that associates with each message transcript $\tau_i$ a cost $c_i(\tau_i) \geq 0$, normalized so that an empty transcript costs $0$.
Once communication is complete, each human player $i$ then selects an action $a_i \in A_i$ as before.  The payoff to each human player is $u_i(\vec{a}, \omega) - c_i(\tau_i)$, their payoff from the baseline game less any costs incurred from communication with the AI agent.  The AI agent receives payoff $v_i(\tau_i, \omega)$.

We emphasize that the payoff to an AI agent can depend on $\tau_i$ and $\omega$, but does not directly depend on realized outcomes or payoffs from the baseline game.  
Consequently, the AI agent's payoff is determined once its interaction with the user is complete.  In essence, a virtual consultant is evaluated on how well it provides advice to its user, rather than how the user chooses to act on that advice.
For example, the AI agent's payoff might depend on the accuracy of information it provides, or even on its user's \emph{anticipated} utility from the baseline game under some behavioral model for the players.  But we do not allow a direct dependence on the user's \emph{realized} utility, as this falls outside the scope of the AI agent's interaction with the user.

\section{Example Scenarios}

{
Before we explore ways that an AI agent's misaligned preferences can impact outcomes, recall that our framework also encompasses settings where an AI agent is simply a technology that reduces costs or increases information by providing signals.  Even such simple settings lead to counter-intuitive outcomes: for example, as with other cost-reducing or signal-inducing technologies, the presence of an AI technology can modify a user's choice set in ways that ultimately reduce the payoffs of the human agent.  See Appendices~\ref{sec:ex-warmup}-\ref{sec:signaling} for concrete examples of this dilemma in the context of an email game.
}

{
That said, our primary focus is on
scenarios where the AI is an agent with preferences that may not be not fully aligned with the user.
We note that while many are concerned with AIs having malicious intent,
misalignment can cause harm even when the AI is essentially benign.%
\footnote{More generally, misalignment can have non-trivial game-theoretic implications that can be either positive or negative. We focus on harms only for ease of presentation.}
The following examples illustrate some of the ways in which our model captures the impact of AI preferences on outcomes.
}

\subsection{Example 1: Advice Evaluated on Perceived Helpfulness}
\label{sec:advice}


{A single user is deciding between a set $A$ of possible actions and can consult an AI agent with access to payoff-relevant information $\omega \in \Omega$.  For instance, the user might be deciding between two similar products and the AI agent knows their features.  The user's preferences are described by a utility function $u \colon \Omega \times A \to \reals$.  A typical interaction may involve the user describing their preferences and the AI agent responding with relevant information or recommendations.
In such a ``chatbot'' scenario, an AI is typically evaluated on (and hence optimized for) metrics like correctness, perceived helpfulness, and relevance.  These goals, which we can think of as an AI's preferences, seem aligned with the user at first glance: after all, helpful, relevant, and personalized advice should lead to better decisions.  But we emphasize that these metrics are functions only of the state of the world and the communication transcript, not the user's realized utility.
Essentially, the AI agent maximizes \emph{perceived helpfulness} at the moment of communication, rather than realized helpfulness in retrospect.
}


While there may be different ways to operationalize ``perceived helpfulness," one can imagine that many of these lead to an AI agent being over-rewarded for being convincing and/or suggesting a new course of action, rather than providing a more balanced view. In particular, the AI may over-emphasize unexpectedly positive or negative features of certain choices, even if they are not fully representative.  Such misalignment may be problematic even if the AI agent is constrained to only return factual information.\footnote{It could also incentivize the AI to hallucinate, but we do not model that in this example.
}

We provide a formal numerical example in Appendix~\ref{sec:advice.example}, demonstrating that this effect can cause a rational user to derive no benefit from the AI agent even though it only ever provides accurate information.
Of course, there may be other ways to set the AI agent's preferences that would avoid this particular issue, {such as an optimization loop in which user satisfaction is judged only in hindsight after actions are taken}. Our point is that the way the AI agent's payoff function is defined can generate unintended distortions by misaligning incentives, and optimizing for a user's declared satisfaction at the moment of interaction with the AI agent could be one such source of distortion.

\subsection{Example 2: Delegated Search}
\label{sec:search}

A human user is attempting to optimize over a space $\Gamma$ of options (think documents or images), and can delegate this task to an AI agent.
The user has a utility function $u : \Gamma \to \reals$ over the space $\Gamma$
all \emph{conceivable} options, from some class $U$ of possible utility functions.  However, only a subset $X \subseteq \Gamma$ of the options are actually
implementable.  {For example, $\Gamma$ might consist of all possible sequences of English words and punctuation, whereas $X$ is a set of grammatically correct and meaningful documents.}
The utility function $u$ and feasible set $X$ constitute the state of nature, drawn from a known prior. We posit information asymmetry: the user knows $u$ but not $X$, and the AI agent knows $X$ but not $u$.\footnote{When we say that the user does not know $X$, what we require is only that they cannot perform optimization over the (possibly very complex) manifold of options.  We allow that the user might recognize an element of $X$ ``when they see it.''} The user can communicate a utility function $u' \in U$ to the AI agent, and the latter returns an option $x \in X$.  The user's payoff is $u(x)$.  The AI agent's payoff is $u'(x) - \gamma(x)$, where $\gamma$ is a fixed penalty function (which can be infinite) chosen by the agent's designer to steer the agent away from some problematic outcomes.

When $X$ is known to the user (i.e., it is drawn from a point mass distribution), this setup defines a Stackelberg game with partially aligned preferences.  Each choice of $u \in U$ has a corresponding maximizer $x(u) \in X$ for the AI agent.  We should expect the human user to report $u' \in U$ that maximizes $u(x(u'))$, yielding the Stackelberg payoff of the game.  That is, the user strategically adapts her
stated preferences,
anticipating the influence of the penalty function.\footnote{{If $\gamma$ and $X$ are perfectly known and the set $U$ includes all possible functions mapping $\Gamma$ to $\reals$, then choosing $u' = u + \gamma$ would always maximize $u(x(u'))$ and, effectively, align the AI with user's incentives. But this generic approach is not necessarily feasible if one or more of these assumptions are relaxed; we describe one such example in Appendix~\ref{sec:search.example}.}}

{More generally, the user only knows a prior distribution over $X$. Then  
the user may still be incentivized to misreport $u$,
just like the full-information case,
but this may now lead to low-utility outcomes on some realizations of $X$.  We provide a numerical example in Appendix~\ref{sec:search.example}, illustrating that attempts to guide an AI agent toward favorable outcomes can ultimately lead to a loss of value when the user is uncertain about the space of available options.}

\section{Research Questions}

We have described how to view AI-based virtual consultants as economic agents and integrate them into some examples of decision- and game-theoretic 
scenarios.
For any given scenario of interest, the resulting economic models can be studied to shed light on key questions about the impact and design of AI agents.  These research questions span both economics and computer science, as outlined below.

\paragraph*{Equilibrium Analysis}  Suppose first that we treat the design of an AI agent as exogenous and fixed.  How does the introduction of the resulting virtual consultants influence strategic behavior and outcomes at equilibrium, both immediately and in the long term?  Do these AI agents increase aggregate welfare for users in the system, and what are the distributive effects?  Which users stand the benefit the most, and which are worse off?  What are the implications for fairness, and are existing biases amplified or suppressed?

While the addition of AI agents can substantially shift the set of equilibria of behavior, there is also the potential to influence the choice of equilibria in games where the equilibrium is not unique.  For example, reliance on AI agents might impact the coordination or correlation of actions of users or shift norms in large populations.  Do cost reductions make it more difficult to build and maintain reputation?  Does the ability to delegate effort to an AI agent make it more or less credible to commit to a given course of action?

\paragraph*{Market Design}  One can also take a more prescriptive approach, and consider the design of virtual consultants and/or platforms to shape incentives and influence outcomes.  The design space for virtual consultants includes both the communication protocol for interfacing with the user, as well as the preference function that guides the AI agent's choice of responses.  Given a game-theoretic scenario of interest, how should these be designed to improve the quality of outcomes when employed by (strategic) human users?  When multiple AI agents are involved, our goal might be to design one of the AI agents fixing
those 
available to other users, or we may be jointly designing all (or some) of the AI agents to achieve good aggregate outcomes.

One can also explore platform and market design under the assumption that users have access to AI agents.  In the context of our general model, this corresponds to designing (some aspects of) the game played between the human users.  As the adoption of virtual consultants becomes more widespread, how does this influence user behavior, and how should platforms and markets adapt?  Should platforms provide their own AI agents, and is there a benefit to designing market rules and AI agents jointly?  What sorts of emergent behaviors might arise from the resulting interactions, and to what extent might optimal market designs make AI agent usage mandatory (or disadvantage those without access)?

\paragraph*{Algorithm design}
Interactions with virtual consultants involve \emph{algorithms}, be they in the implementation of the AI agents themselves or in the way they are used.  Implicit in all of our models is the question of optimizing one's communication with an AI agent, whether directly via algorithmic layer or indirectly through human user behavior.  Likewise, an AI agent faces an optimization problem when generating responses, which can be guided by explicit algorithmic architectures built on top of LLMs.  Which algorithm designs, on either side, lead to more desirable outcomes?  How should performance be evaluated, and how should we value robustness and stability versus expected performance?  How should the cost of using an AI agent be modeled, and how can we optimize performance-cost tradeoffs?

\section{Extending the General Model}

We presented a model that aims to capture the introduction of AI-powered virtual consultants as economic agents with limited agency.  There are natural ways to extend this perspective to encompass other economic questions related to AI agents.  Here we briefly describe a few high-level questions that one might extend our general model to include.

\paragraph*{Increased AI Agency}  We intentionally modeled AI agents as consultants with a limited degree of agency.  In our model, AI agents can send messages to their human users but cannot (a) interact directly with other virtual consultants or other human users, or (b) directly take actions that are payoff-relevant for human users.  As AI technology evolves, we may see scenarios in which AI agents become empowered to do one or both of these.  For example, one might imagine an AI agent for resume screening that is empowered to reject candidates without human input in certain cases.  From a modeling perspective, one might still attempt to model such a scenario as a human-taken action, implemented through a sign-off policy that the human user ultimately has control over.  However, if delegating actions to AI agents becomes increasingly more commonplace, we may reach a point where it is more natural to endow agents with increased (but still limited) agency within our models.

\paragraph*{Platforms and Interfaces for AI Agents}  Our model treats an AI agent's information set and internal optimization as intrinsic to the agent.  However, one might explicitly model the manner in which an agent acquires external data or otherwise interacts with the world, especially if one also increases the agency of an AI agent.  For example, an AI agent that provides recommendations for jobs to apply to (or one that helps screen job candidates) might benefit from interacting with a job-search platform.
But if such agents become commonplace, it is natural to imagine that a platform could be incentivized to offer data interfaces that are tailored to
virtual consultants.  More broadly, one can imagine ecosystems for services and/or data to be used by virtual consultants, enabling general-purpose personal AI assistants that can serve as interfaces with the broader world through one's smartphone.  And indeed some nascent markets for virtual consultant services are already emerging.%
\footnote{E.g., plugins for use with OpenAI's ChatGPT.}  Modeling the ways in which virtual agents can interact with platforms, and/or how any platform usage costs would be internalized by agents (or otherwise how platforms might monetize such interactions), is likely to be of growing relevance as these ecosystems mature.

\paragraph*{Economics of Training AI Agents}  Our modeling to this point treats AI agents as entities in a scenario, but one might additionally wish to model the manner in which such agents are created and/or maintained.  This might involve modeling the data used to train ML models that drive the AI agents, the way that an AI agent changes in response to its training data, and the firm(s) that provide these AI agents to users.  By modeling the source of the data used to train AI agents (which may come directly from the real world, or may be generated by another AI), one can explore the economics of data generation and monetization. How are content creators incentivized to
contribute to the training data,
and how does this influence the content they create?  How are the distribution platforms or other repositories on which that data resides incentivized to monetize and/or protect that data, and how does this impact the data on which AI agents are trained?  What influence might this have on the behavior, information sets, or (implicit) incentives of AI agents?

\section{Related Literatures}

Our framework is related to several literatures in economics and computer science. We highlight these connections (and outline the differences) in what follows.%
\footnote{Throughout, we only provide brief, exemplary citations, opting for surveys whenever possible. A more detailed survey of these literatures is beyond our scope.}

\paragraph*{Delegation and Contracting} Our central argument is that modern AI assistance technologies are best modeled as economic agents.  As such, our framework is closely related to principal-agent models in which a principal decision maker delegates to another (human) agent.  This includes the theory of delegation and of contract design, both well-studied in microeconomics (see, e.g.,~\cite[Chapter~14]{mas1995microeconomic} and \cite{LM02} for an introduction to principal-agent models,~\cite{bendor2001theories} for an overview of the theory of delegation, and~\cite{holmstrom1984theory,armstrong2010model} for other representative delegation models) and increasingly in economics and computation.

The theory of delegation focuses largely on
consequences of misaligned incentives,
where the principal and agent may prefer different actions be taken.  The principal can attempt to mitigate this issue by committing to the way in which they will act on advice, or restricting the space of options available to the agent.  A crucial aspect of such models is that the agent has preferences over the choice of action that is ultimately taken.
In contrast, we model an AI agent as having preferences over
{the communication transcript itself, rather than actions or realized outcomes.  This might mean, for example, that the agent is indifferent between aspects of the principal's preferences that are not revealed during communication, even if they would influence the principal's choice of action.}
In this sense our modeling of AI agent incentives is more closely related to contract design, in which agents commonly have preferences driven by the cost of taking actions but not over the outcome realizations.
One difference between our model and contract design
is that we assume AI agents cannot be incentivized with monetary transfers;
rather, they are internally incentivized with preferences over the communication process, as mentioned above.

\paragraph*{Algorithmic Interfaces and Human-AI Collaboration} It is increasingly common for online platforms and markets to support AI-powered interfaces and algorithms that offload platform-specific actions.  These include autobidding services popularized in advertising platforms~\cite{GoogleAutobidder,BingAutobidder,FacebookAutobidder}, price prediction and recommendation services for matching platforms like AirBnB and Amazon~\cite{AirbnbPricing,AmazonPricing}, high-frequency trading algorithms~\cite{o2015high}, and more.
There is a substantial and growing literature on the design and use of such systems, and how they ultimately impact the behavior of (human) users.  Of particular interest is the goal of automating the behavior of the users themselves, as a way of performing those tasks more quickly, cheaply, and/or responsively than a human can.  Recent work has begun to explore the use of generative AI in such scenarios, including automating the generation of content, and implications for platform design~\cite{duetting2023mechanism,yao2024human,fish2024algorithmic}. Relative to these scenarios, the AI tools we seek to model are less tied to the structure of a specific domain or platform and, crucially, do not directly take actions on their users' behalf.

The design of AI-powered systems and interfaces is heavily informed by the theory of Human-AI collaboration and user behavior.  This literature focuses on human-in-the-loop systems, analyzing the performance of AI systems in the context of (and in anticipation of) how they will be used~\cite{grosz1996collaborative,amershi2019guidelines}.  This includes analyzing the importance of features like trust~\cite{hoff2015trust}, 
interpretability~\cite{gilpin2018explaining}, fairness~\cite{mehrabi2021survey}, and predictability~\cite{bansal2019updates}. Our general model abstracts away from specific application domains and interfaces, but as with the study of Human-AI collaboration it is crucial that a human user understand the AI's communication protocol and preferences in order to correctly and usefully interpret any given AI interaction.

\paragraph*{Information Design and Decision-Making under Imperfect Information}  We model AI agents as virtual consultants that provide information to human users.  This is closely related to the impact of changing the amount of information available to a human agent when taking actions or making decisions.  For instance, there is a growing line of literature on how to use historical sales data to set prices and/or design the rules of a marketplace or auction~\cite{den2015dynamic,cole2014sample,morgenstern2016learning}, how buyer and/or seller behavior is impacted by the quantity or quality of data they have access to~\cite{bergemann2015limits,bergemann2019information}, and how mechanism design can be augmented by imperfect predictions~\cite{agrawal2022learning,xu2022mechanism}.  Our model includes such scenarios by assuming that the data is provided by and accessed via an AI agent.  The primary difference is that we view an AI agent as having some degree of agency and preference, so that rather than simply passing unfiltered data the AI agent may provide some additional processing or analysis on the user's behalf.

The notion of filtering data evokes the study of information design~\cite{BergemannMorris-infoDesign16}, in which the ``designer"
strategically selects which information to provide to players in a game.  In Bayesian Persuasion, for example, a sender commits to a protocol that determines which (partial) signals to reveal to a receiver, in order to induce a desirable behavior
~\cite{kamenica2011bayesian,Kamenica-survey19}.  Our model differs in that the AI agent does not have a stake in the action ultimately taken by its human user, so there can be no temptation to persuade.  However, the AI agent may still have (possibly misaligned) incentives over the communication transcript and this can influence the nature of information that is shared.  Analyzing the outcome of communication with AI agents therefore shares much in common with analyzing the space of behaviors that can be obtained under different information structures.

\paragraph*{Simulating Strategic Agents with AI}
There is a recent and rapidly growing line of research using large language models to explicitly simulate the behavior of strategic agents.  This includes simulating the outcome of game-theoretic studies in economics~\cite{horton2023large,shapira2024can}, generating virtual panels of AI-powered individuals for market research~\cite{brand2023using}, or designing AI agents to play strategic games such as Werewolf or Diplomacy~\cite{xu2023exploring,meta2022human}.  These studies suggest that modern LLMs have a (perhaps implicit) capacity for strategic reasoning, encoded within their training data, that can be measured and evaluated~\cite{raman2024rationality}.  In a similar vein, a line of work on ``common sense reasoning'' by AI systems has demonstrated consistent improvement by LLMs on challenge datasets that are designed to encode and test ``knowledge that is commonly assumed in other humans"~\cite{davis2015commonsense,sakaguchi2021winogrande,li2023textbooks,touvron2023llama}.  Again, this suggests
a notion of common-sense reasoning embedded in LLMs that could be distilled and recovered.

Together, these lines of work support the idea that AI-powered agents can have latent knowledge and behaviors that resemble human economic agents.  But we do not restrict our attention to scenarios where an AI system is explicitly directed to simulate a human or an economic agent.  Rather, a modern AI implicitly acts as an economic agent for \emph{all} tasks, including tasks that are not directly economic or strategic.  The fact that generative AI can effectively simulate
human-like common sense and strategic reasoning
supports
our choice to model them as economic agents.

\paragraph*{Designing and Evaluating Generative AI}
Underlying the economic analyses of generative AI tools, there is a huge and rapidly growing literature on generative AI itself, and particularly LLMs.
The scope is very broad: from designing the AIs (including foundation models and training thereof \cite{transformer-neurips17,gpt3-neurips20,gpt4-tr23,Llama2-2023,Phi1-arxiv23} and fine-tuning \cite{RLHF-neurips22,RLHF-anthropic22}; see
\cite{LLM-survey-zhao23} for a survey) to orchestrating their usage (e.g., prompting techniques \cite{gpt3-neurips20,CoT-Wei22,CoT-Kojima22}, integration with tools \cite{Wolfram-gpt23,gao2023pal}) to evaluating their capabilities \cite{Sparks-bubeck23,LLM-eval-survey24}. While economic models are abstract by nature and hide most implementation details, the models (and the questions being asked) should be grounded in the state-of-art in machine learning: which capabilities are feasible now, what is projected to be realistic (or not) in the near future, and what are the salient costs and tradeoffs. This is particularly important when economic models need to incorporate some domain-specific structures, such as costs or constraints, that may be first-order concerns in a given application scenario.

%% file: main-appendix.tex
\section{Detailed Examples}

\subsection{Warm-up: Single-Player Email Game}
\label{sec:ex-warmup}

The following simple scenario is intended as a warm-up
and showcases how our framework incorporates delegation.
A single human player is writing an email and is debating between two different styles, $A$ or $B$.  The state of nature $\omega \in \{A,B\}$ determines which email style is most appropriate for the given scenario.  The human's prior is that the state is equally likely to be $A$ or $B$.  The human can choose whether to think carefully about the situation, which comes at a cost of $4$ but reveals the state of nature.  They then choose whether to write email $A$ or $B$.  Choosing the correct style generates value $5$ while choosing the incorrect style generates value $-10$.

Imagine now that the human player has access to an AI agent.  The AI agent has access to a signal $\sigma \in \{A,B\}$ that is equal to the true state with probability 90\%.
\footnote{Formally, the state of nature is now a pair $(\omega, \sigma) \in \{A,B\}^2$,  initially drawn from a joint distribution.}
The human player can choose whether to request a signal $m \in \{A,B\}$ from the AI agent.  Doing so incurs a cost of $1$ for the user.  The AI obtains utility $1$ if it correctly reports the state of nature and $0$ otherwise.  Given this utility and information structure, the AI will maximize utility by reporting $m = \sigma$. In this case, the human maximizes utility by requesting a signal $m$ from the AI, choosing not to think carefully, then writing an email in style $m$.  This results in an expected payoff of $(0.9)(5) + (0.1)(-10) - 1 = 2.5$ for the user.  Notably, this results in a reduced probability of selecting the appropriate email style.


\subsection{Costly Signaling}
\label{sec:signaling}


A professor is writing a recommendation letter for a student.  The student is equally likely to be either Typical ($T$) or Strong ($S$), and this is encoded in the state of nature $\omega \in \{T,S\}$.  There are two human players, a sender (the professor) and a receiver.  The sender observes the student's strength, but the receiver does not.  The sender first chooses whether to write a weak or strong letter.  Writing a weak letter is free but writing a strong letter costs the sender $4$.  The receiver then observes the sender's letter and chooses whether to hire the student.
The receiver obtains payoff $1$ for hiring a strong student, payoff $-2$ for hiring a typical student, and payoff $0$ for not hiring.  The sender obtains value $10$ if a strong student is hired, value $6$ if a typical student is hired, and value $0$ if the student is not hired.\footnote{One could obtain a qualitatively similar outcome if the sender's payoff if the student is hired is independent of quality, but writing a strong letter is easier/cheaper for strong students.}

This game has two equilibria.  First, there is a trivial ``babbling equilibrium" where the receiver never hires (i.e., ignores the letter) and the sender always writes a weak letter.  But there is also a non-trivial equilibrium, which we will describe first in terms of the receiver's strategy and then the sender's strategy.
The receiver does not hire upon receiving a weak letter, and hires with probability $p$ upon receiving a strong letter.
The sender always writes a strong letter for strong candidates, and writes a strong letter for a typical candidate with probability $q$.
We have an equilibrium at
$(p,q) = \rbr{\nicefrac{2}{3}, \nicefrac{1}{2}}$.
Indeed, if $q = \nicefrac{1}{2}$ then, conditional on receiving a strong letter, the receiver's posterior belief is that the student is strong with probability $\nicefrac{2}{3}$, which makes her indifferent between hiring and not hiring.  If $p = \nicefrac{2}{3}$ then the sender strictly prefers to write a strong letter for strong students (expected total payoff $10 \times \nicefrac{2}{3} - 4$ for a strong letter versus payoff $0$ for a weak letter), and is indifferent between writing strong and typical letters for typical students (expected total payoff $6 \times \nicefrac{2}{3} - 4 = 0$ for a strong letter).
Under this equilibrium, a strong student is hired with probability $\nicefrac{2}{3}$ and a typical student is hired with probability $\nicefrac{1}{2} \times \nicefrac{2}{3} = \nicefrac{1}{3}$.
The sender's expected utility is $\nicefrac{1}{2}\rbr{10 \times \nicefrac{2}{3} - 4} = \nicefrac{4}{3}$, the surplus obtained from writing strong letters for strong students.

Next imagine that the sender has an AI agent that can assist with the letter-writing process, so that the cost of writing a strong letter effectively reduces from 4 to 1.  (The receiver's AI agent, if any, does not provide any assistance in this matter.)  If we repeat the equilibrium calculation above, our choice of $q=\nicefrac{1}{2}$ will be unchanged, but making the sender indifferent between strong and weak letters for typical students requires that the receiver considers the letter with probability only $p = \nicefrac{1}{6}$.  Thus, at equilibrium, a strong student is hired with probability $\nicefrac{1}{6}$ and a typical student is hired with probability
$\nicefrac{1}{2} \times \nicefrac{1}{6} = \nicefrac{1}{12}$.  In this equilibrium the sender's utility falls to
$\nicefrac{1}{2}\rbr{10 \times \nicefrac{1}{6} - 1} = \nicefrac{1}{3}$, which is again the surplus obtained from writing strong letters for strong students. Notably, not only are students hired less often, but the sender's expected utility falls at equilibrium due to the presence of the AI agent. This is because the sender cannot commit to \emph{not} using the AI agent, and this limits the sender's ability to signal the student's type via costly effort.

\subsection{Advice Evaluated on Perceived Helpfulness: Numerical Example}
\label{sec:advice.example}


The following is an instance of the AI advice game described in Section~\ref{sec:advice}.
A user is deciding which of two products to purchase and can consult an AI agent. Product A is a safe choice that always results in utility $0$ for the user.
Product B has a two binary features, 
$\omega = (\omega_1, \omega_2) \in \{0,1\}^2$, that are drawn uniformly at random.
The realization of $\omega$ is known to the AI agent but not the human user.  The user has a utility function
$v \colon \{0,1\}^2 \to \reals$
that maps realizations of $\omega$ into a (possibly negative) payoff for purchasing Product B.  The user can communicate with the AI agent before making a purchasing decision.
Specifically, if the user requests advice, the AI sends a message that contains one (and only one) of the product features.\footnote{The restriction to declaring only one feature
is motivated by communication length; in general an AI agent cannot practically communicate all information it has access to.}

Suppose that, motivated by a design objective of maximizing perceived helpfulness, the AI agent's payoffs are determined as follows.  Declaring a feature incorrectly gives payoff $-\infty$ for the AI.  Otherwise, the AI agent's payoff is the difference in expected utility between the user's optimal choice given the revealed feature(s) and their optimal choice given no information, where the expectation is over any features that are not revealed. This payoff is always non-negative, and is $0$ whenever the AI agent's advice would not influence a utility-maximizing user's choice of action.




Suppose that the user's utility function is $v(1,1) = 2$ and $v(0,0) = v(0,1) = v(1,0) = -1$.  I.e., the user gets utility $2$ for buying B if both features are $1$, otherwise utility $-1$.  In this case, absent any information from the AI agent, the expected utility-maximizing action is to purchase A since $\nicefrac{3}{4}\cdot(-1) + \nicefrac{1}{4}\cdot 2 < 0$.  If either feature is revealed to be $0$, then the user's utility is certainly $-1$ for purchasing B so the optimal choice is still purchase A.  But if either feature is revealed to be $1$, the expected utility from purchasing B (in expectation over the other feature) is $\nicefrac{1}{2}\cdot(-1) + \nicefrac{1}{2}\cdot 2 = \nicefrac{1}{2} > 0$.  This means that correctly revealing a feature to be $1$ yields payoff $\nicefrac{1}{2}$ for the AI, whereas revealing a feature to be $0$ yields payoff $0$.  We emphasize that this payoff to the AI is independent of which product the user ultimately decides to purchase; it is a function only of $\omega$ and the communication transcript.

This situation creates misaligned incentives for the human user and the AI.  If the realized state is $\omega = (1,0)$, then the user's ex post preferred action is to purchase A.  But under this realization the AI is incentivized to reveal $\omega_1 = 1$.  In fact, the AI is incentivized to reveal a feature with value $1$ in all states except $\omega = (0,0)$.  What this means is that, conditional on the AI revealing a feature $\omega_i = 1$, a rational user should infer only that $\omega \neq (0,0)$, so her expected utility from purchasing B is $\nicefrac{2}{3}\cdot(-1) + \nicefrac{1}{3}\cdot 2 = 0$.  In other words, the AI agent confers zero value even to a fully rational user that correctly anticipates its behavior.

\subsection{Delegated Search: Numerical Example}
\label{sec:search.example}

{
The following is a toy example of the AI delegated search game described in Section~\ref{sec:search}.  The example illustrates a scenario where (a) a user can potentially benefit by strategically inflating the strength of their preferences in order to bypass the AI agent's preference to avoid certain outcome, but (b) this can lead to suboptimal outcomes when the set of possible outcomes available to the AI agent is not perfectly understood by the user.
}

{
In this toy example, a user is writing an email with the assistance of an AI agent.  Emails are evaluated on two dimensions, say humor and professionalism.  Under this parameterization, $\Gamma = [0,1]^2$ denotes the space of all conceivable emails, where $(1,0)$ is a perfectly humorous but unprofessional email and $(0,1)$ is perfectly professional but not at all funny.  Not all points in the parameter space are necessarily achievable, so while $(1,1)$ represents a hypothetical email that is both perfectly funny and perfectly professional, this may not be possible to achieve.  Here $X \subseteq [0,1]^2$ is the set of all achievable parameters.
}

{
The user's utility from an email is characterized by a relative preference for humor versus professionalism.  This is described by a vector $y \in \reals_{\geq 0}^2$ with $||y||_2 = 1$.  An email $x \in \Gamma$ then generates utility $u_y(x) = x \cdot y$, where we write $u_y$ for the utility function corresponding to preference vector $y$.  We let $U$ be the space of all such utility functions.
}

{
Let's say the user knows that $X$ contains $x^{\text{f}} = (1,0)$ and $x^{\text{p}} = (0,1)$ for ``funny" and ``professional", respectively,
so emails achieving these parameters exist.
Let's further imagine that the AI agent is designed to downplay humor unless specifically requested; this is encoded as a penalty function with $\gamma(x^{\text{f}}) = 0.2$ and $\gamma(x^{\text{p}}) = 0$.  One can then verify that if $X = \{x^{\text{f}}, x^{\text{p}}\}$, the AI agent will return $x^{\text{p}}$ in response to declared relative preferences $y = (y_1,y_2)$ if $y_1 < 0.8$ (and hence $y_2 > 0.6$), and will return $x^{\text{f}}$ if $y_1 > 0.8$.  Thus, if the user has a strict preference for the humorous email (no matter how slight), they will optimize by declaring a strong preference for humor.
}

{
Next imagine that the AI agent is also able to create emails that combine professionalism and humor to a certain extent.  Concretely, there are points $x^{\text{m}} = (\nicefrac{2}{3},\nicefrac{2}{3})$ (`m' for mediocre at both) and $x^{\text{pf}} = (\nicefrac{1}{2},1)$ (`pf' for professional yet funny) in the set $X$ as well.  But the AI agent steers away from including humor in professional emails; say this is encoded with penalties $\gamma(x^{\text{m}}) = 0.25$ and $\gamma(x^{\text{pf}}) = 0.35$.  These numbers are chosen so that the threshold on the declared preference for humor at which
$x^{\text{m}}$
is returned, versus
$x^{\text{pf}}$,
is $\approx 0.74 < 0.8$.
}

{
A feature of this example is that if the agent's most-preferred option is
$x^{\text{pf}}$,\footnote{For example, this is the case if the agent's true preference is given by $z = (z_1,z_2)$ with $z_1 = 0.79$.} then it would be optimal to declare a moderate preference for humor; say $y_1 = 0.71$.  On the other hand, if the user is unaware that this option is possible and prefers
$x^{\text{f}}$ to $x^{\text{p}}$,\footnote{Which is likewise obtained by true preferences $z = (z_1,z_2)$ where $z_1 = 0.79$} they would rationally declare a stronger preference for humor, $y_1 > 0.8$, which would be suboptimal with respect to the true choice set.  This toy example demonstrates that while an agent can benefit by strategically declaring their preferences to an AI that attempts to steer away from certain outcomes, this can lead to suboptimal outcomes when the set of possible options is not perfectly understood. While this example is described in terms of an incorrect belief about $X$, the same conclusion can be reached if the user has a prior over $X$, in which case a suboptimal choice is reached on some realizations of $X$.
}